\documentclass{aa}  
\usepackage{graphicx}
\newcommand{\Msolar}{\mbox{$M_{\odot}\,$}}

\def\gs{\mathrel{\raise0.35ex\hbox{$\scriptstyle >$}\kern-0.6em \lower0.40ex\hbox{{$\scriptstyle \sim$}}}}
\def\ls{\mathrel{\raise0.35ex\hbox{$\scriptstyle <$}\kern-0.6em \lower0.40ex\hbox{{$\scriptstyle \sim$}}}}
\newcommand{\arcsecs}{\mbox{$^{\prime\prime}$}}

\begin{document}
   \title{CO Line Emission from Lyman Break Galaxies}

   \subtitle{Cosmological Simulations and Predictions for ALMA}

   \author{T.R. Greve
          \inst{1,2}
          \and
          J. Sommer-Larsen\inst{3}%\fnmsep
          }

   \offprints{T.R. Greve}

   \institute{Max-Planck-Institut f\"{u}r Astronomie,
              K\"{o}nigstuhl 17, D-69117 Heidelberg, Germany\\
              \email{tgreve@mpia-hd.mpg.de}
         \and
              California Institute of Technology,
              1200E California Blvd., CA 91125, USA
         \and
             Dark Cosmology Centre, 
             Juliane Maries Vej 30, DK-2100 Copenhagen {\O}, Denmark\\
             \email{jslarsen@tac.dk}
             }

   %\date{Received September 15, 1996; accepted March 16, 1997}
   \date{}

% \abstract{}{}{}{}{} 
% 5 {} token are mandatory
 
  \abstract
  % context heading (optional)
  % {} leave it empty if necessary  
   {The detection of the rotational lines of CO in proto-galaxies in the early Universe
   provides one of the most promising ways of probing the fundamental physical properties
   of a galaxy, such as its size, dynamical mass, gas density, and temperature.
   While such observations are currently limited to the most luminous galaxies, the advent of ALMA
   will change the situation dramatically, resulting in the detection of numerous normal galaxies
   at high redshifts.
   }
  % aims heading (mandatory)
   {We investigate the morphology and strength of the CO rotational line emission emerging from $z\sim 3$ 
   progenitor systems of normal field galaxies seen in the present-day Universe, 
   examine how well ALMA will be able to detect such systems and how accurately 
   CO line widths, gas morphologies and ultimately dynamical masses can be inferred.
   }
  % methods heading (mandatory)
   {Maps and spectra of rotational CO line emission were calculated
    from a cosmological N-body/hydrodynamical TreeSPH simulation of a $z\sim 3$ 
    "Lyman break galaxy" of UV luminosity about one order of magnitude below $L^*$. 
    To simulate a typical observation of our system with ALMA, we imposed characteristic 
    noise, angular, and spectral resolution constraints.  
    }
  % results heading (mandatory)
   {
    The CO line properties predicted by our simulation are in good
    agreement with the two Lyman break systems detected in CO to date.
    We find that while supernovae explosions from the ongoing star formation 
    carve out large cavities in the molecular ISM,
    they do not generate large enough gas outflows to make a substantial imprint on 
    the CO line profile. This implies that for most proto-galaxies 
    -- except possibly the most extreme cases -- stellar feedback effects 
    do not affect CO as a dynamical mass tracer. 
   }
  % conclusions heading (optional), leave it empty if necessary 
   {
    Detecting CO in sub-$L^*$ galaxies at $z\simeq 3$ will push ALMA to the limits of its cababilities,
    and whether a source is detected or not may depend critically on its inclination angle.
    Both these effects (sensitivity and inclination) will severely impair the ability of ALMA 
    to infer the gas kinematics and dynamical masses using line observations.
    }

   \keywords{cosmology -- theory -- galaxy formation -- ISM -- molecules -- numerical
               }

   \maketitle
%
%________________________________________________________________

\section{Introduction}
Since the first discovery of CO in IRAS\,F10214$+$4724 at $z=2.29$ 
(Brown \& Vanden Bout 1992), observations tracking the molecular gas 
(H$_2$) in distant galaxies have taken on 
an increasingly important role as a way to
study galaxy formation and evolution (Solomon \& Vanden Bout 2005).
Prime reasons for this are that CO provides an 
estimate of the total molecular gas reservoir available for
star formation and black hole accretion and can be used as a  
tracer of the dynamical mass of a galaxy (Bryant \& Scoville 1996). 
The latter is particularly important during the formative stages of
a galaxy, where extreme dust-obscuration, as well as outflows, can make
dynamical mass estimates derived from optical/near-IR spectroscopy 
unreliable. 

Today, only the most extreme objects at high-$z$, such as QSOs and 
luminous submillimeter-selected galaxies (SMGs -- Blain et al.\ 2002), are detectable in
CO. In contrast, the more numerous, but also less massive and presumably less gas-rich, 
Lyman break galaxies  (LBGs -- Giavalisco 2002) are generally too faint to be detected in CO
with current instruments (cf.\ Baker et al.\ 2004a).
With the advent of the Atacama Large Millimeter Array (ALMA), this situation may change, making
it important that realistic numerical models are developed 
in order to interpret the data. 
In spite of this, the amount of numerical/theoretical work on simulating the expected 
CO emission from primordial galaxies has been limited to a few pioneering studies
(Silk \& Spaans 1997; Combes, Maoli \& Omont 1999).
While these initial efforts were mainly focused on the detectability of high-$z$ CO lines
with future mm-facilities, using 'idealized' generic starburst galaxies as templates,
more recently, however, Narayanan et al.\ (2006) simulated
the CO emission from two merging gas-rich disk galaxies in detail,
and studied the effects of a massive black hole on the gas dynamics
of the system as revealed by the CO line emission.\\
\indent In this letter we present CO intensity maps and line profiles
of a cosmological self-consistent simulation of a merger system,
representative of the LBG population at $z\sim 3$.
We explore the feedback effects from supernovae (SNe) on the molecular gas, and in particular whether
they affect dynamical mass estimates based on CO. 
The simulation is subjected to the sensitivity and resolution
constraints imposed by ALMA, and we explore the effects this
have on our results.
We adopt a flat cosmology with $\Omega_m=0.27$, $\Omega_\Lambda=0.73$ and
$H_0=71$\,km\,${\mbox{s}^{-1}}$\,Mpc$^{-1}$ (Spergel et al.\ 2003).

\section{Simulations}\label{section:model}
The cosmological simulation of the formation and evolution of an individual
galaxy was performed using the N-body/hydrodynamical 
TreeSPH code, briefly described in Sommer-Larsen (2006).
The system becomes a Milky Way/M31 like disk galaxy at $z$=0, and is
simulated using about 2.2 million particles in total, comprising only
SPH and dark matter (DM) particles at the initial redshift, $z_i$=39. 
The masses and gravity softening lengths
of SPH and star particles are $9.9\times 10^4\,h^{-1}$M$_{\odot}$ and
$200\,h^{-1}$pc, respectively. For the DM particles the corresponding
values are $5.7\times 10^5\,h^{-1}$M$_{\odot}$ and $370\,h^{-1}$pc. The minimum
SPH smoothing length in the simulation is about $13\,h^{-1}$pc.
For the purposes of this paper, a simulation output at $z=3.26$,
was chosen. We emphasize, that the merger-nature and physical properties 
of the proto-galaxy at this time are derived self-consistently 
from a cosmological simulation, invoking detailed baryonic physics,
such as radiative gas cooling, star formation, chemical enrichment
and feedback from SNe. Note, however, that only atomic
radiative cooling is invoked, hence neutral, high-density gas does not
cool much below $T\sim$10$^3$ K, and ``sub-grid'' modeling of the molecular
gas phases are required (see below). Moreover, AGN feedback is not
included in the simulations; this may play a role in the formation of
early type galaxies, but likely not in late types (e.g.\ Sommer-Larsen 
2006).

At $z=3.26$, the system has already undergone significant chemical enrichment, and a substantial
fraction of the gas has assembled into a number of starforming disk systems, some of which are in the
process of merging. Here, we focus our attention on the central 6\,kpc of the region, which
contains the main disk system (A), and a 
second, somewhat smaller system (B) which is undergoing a merger with A. 
Both systems have a combined star formation rate of $\sim 30\,\Msolar\,\mbox{yr}^{-1}$, 
corresponding to a UV luminosity about one magnitude below $L^*$ (Sommer-Larsen \& Fynbo 2007),
The stellar masses are $M_{stars}\simeq 6.5\times 10^9\,\Msolar$ (A)
and $M_{stars}\simeq 3.2\times 10^9\,\Msolar$ (B), which is
within the range typically observed for LBGs at $2\ls z\ls 3$ (Shapley et al.\ 2005).

Incorporating the formation of molecular gas (H$_2$) from the atomic
phase (H\,{\sc i}) into numerical models in a fully self-consistent manner is
currently intractable -- largely owing to the lack of computational
power, and our incomplete understanding of how H$_2$ forms in the
first place. As a result, we adopted physically motivated recipes for the formation of H$_2$ from the
atomic phase. These included the expected scaling of the H$_2$ formation rate with the metallicity ($Z$)
and pressure ($P\propto n(T + \sigma_v^2)$) of the gas (Elmegreen 1989, 1993), as well as the 
increase in the H$_2$ destruction rate with increasing ambient UV field ($G_o$, in units of
an average interstellar flux between $6\,\mbox{eV}\ls h \nu \ls 13.6\,\mbox{eV}$ of $1.6\times 10^{-3}\,\mbox{erg\,cm$^{-2}$\,s$^{-1}$}$
 -- Habing 1968). We ignored, however, the destruction of H$_2$ due to cosmic rays (Cazaux \& Spaans 2004).
Also, we made the simplifying assumption that the formation and destruction 
of H$_2$ is in equilibrium. In this case, the local molecular gas mass fraction
can be expressed analytically as a function of $Z$, $P$, and $G_o$ (Pelupessy, Papadopoulos \& van der Werf 2006), 
which we then used to derive the distribution of H$_2$ in our simulation. 
Within the starforming disks in our simulation, a fixed $G_o = 10^5$ was adopted, typical of
irradiated gas near an OB-association. Away from the disks, the average value of the local interstellar
radiation field was used ($G_o=1.7$ -- Draine \& Salpeter 1978). 
The macroscopic velocity dispersion, $\sigma_v$, of the H\,{\sc i} gas was calculated
from the velocities of the SPH particles. We find component A to have molecular and atomic gas masses of 
$M(\mbox{H}_2) \simeq 1.4\times 10^9\,\Msolar$ and $M(\mbox{H\,{\sc i}})\simeq 4.0\times 10^8\,\Msolar$, respectively,
while for component B we find 
$M(\mbox{H}_2) \simeq 8.0\times 10^8\,\Msolar$ and $M(\mbox{H\,{\sc i}})\simeq 1.8\times 10^8\,\Msolar$.
Thus, most of the gas within the disks are in the molecular phase, consistent with what is 
seen in local galaxies, where most of the neutral gas resides on much larger scales ($\gs 10\,$kpc).
The H$_2$-to-H\,{\sc i} mass ratios are within the range observed for luminous starburst galaxies (Mirabel \& Sanders 1989).
In order to resolve scales comparable to that of Molecular Clouds Complexes, which in our
Galaxy are found to have typical size of $\sim 50-100$\,pc (Stark \& Blitz 1978), 
a grid cell size of 10\,pc was adopted. Using 
5 and 20\,pc cells did not change our findings in any significant way.
The H$_2$ mass within each cell was assumed to reside in
a single cloud, following the same large-scale motion within the galaxy as the atomic gas,
and obeying the same correlations between velocity dispersion and size/mass as observed
for a wide range of Galactic clouds (Larson 1981). This then allowed us to determine
the H$_2$ density as well as the area filling factor of each cloud across the
grid. The range in H$_2$ densities obtained in this way was $\sim 50 - 2\times 10^6\,\mbox{cm}^{-3}$,
which closely matches the densities typical of the molecular gas responsible for 
CO line emission in our own Galaxy.

A Large Velocity Gradient (LVG) model (e.g.\ Goldreich \& Kwan 1974) was employed
to calculate the CO population levels and source functions for each cloud, using its H$_2$ density
together with fixed values of the kinetic temperature and CO abundance 
([CO/H$_2$]$=10^{-4}$; the C and O abundance of the cold gas in the two central disks is approximately solar). 
Guided by studies of local starburst galaxies, which have been shown 
to harbor a two-phase molecular ISM (Aalto et al.\ 1995),
clouds with a mean density $\gs 10^4\,\mbox{cm}^{-3}$ were given a constant kinetic temperature of $T_k=55\,$K, while
clouds with lower densities were assigned a temperature of $T_k=80\,$K.
We used a Cosmic Microwave Background temperature of 
$T_{\mbox{\tiny{CMB}}} = (1+z)\times 2.73\,\mbox{K} \simeq 11.63\,$K, although the
effects of the CMB radiation on the CO excitation are negligible, except for very cold gas.
The final emerging CO emission was then obtained by integrating the radiative transfer equation
along each line of sight, taking into account the area filling factor of each cloud.

\section{Results \& Discussion}\label{section:results}
\subsection{Intensity maps and line profiles} 
\begin{figure*}
   \centering
   \includegraphics[width=0.500\hsize]{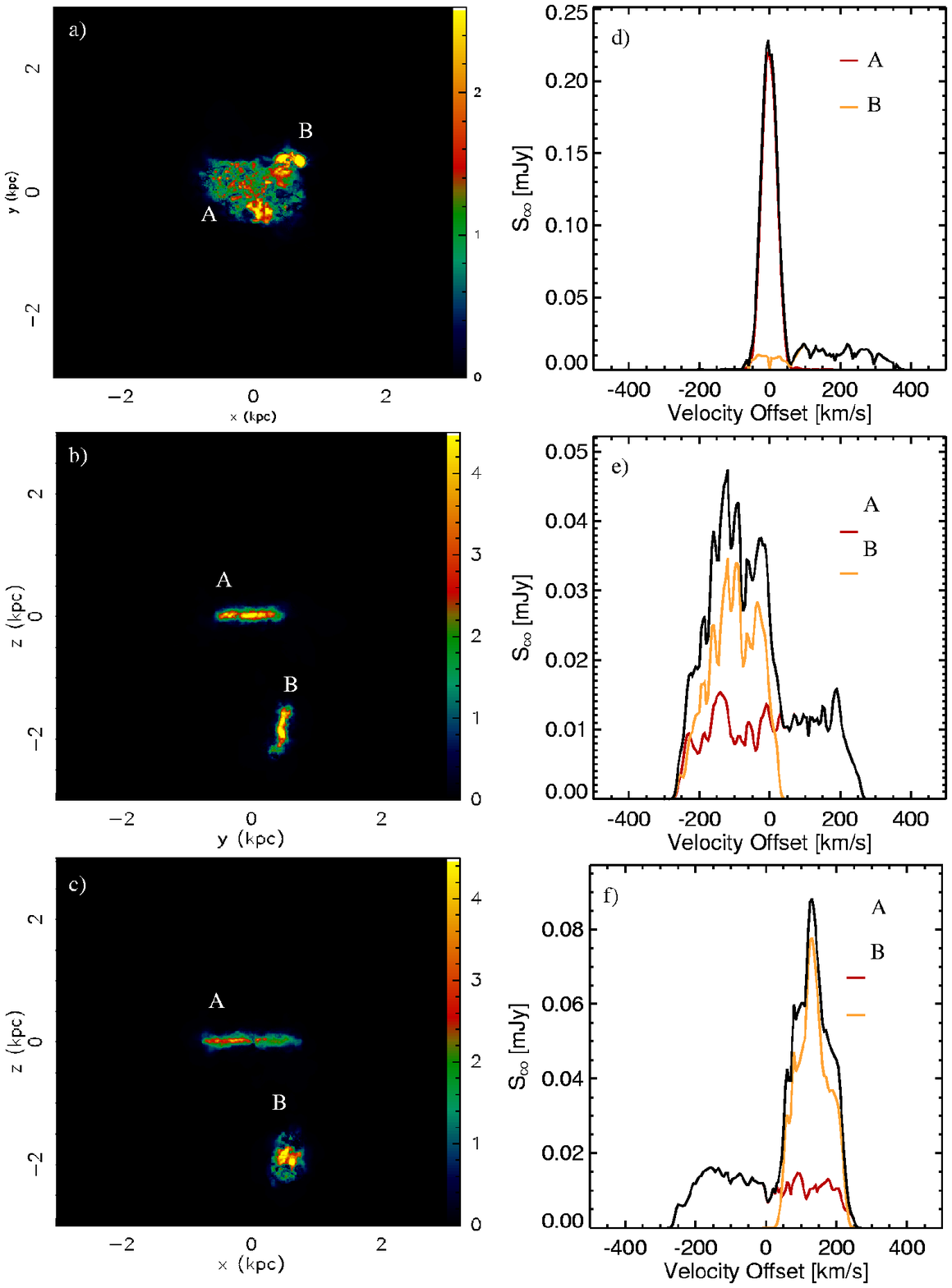}~~~~\includegraphics[width=0.475\hsize]{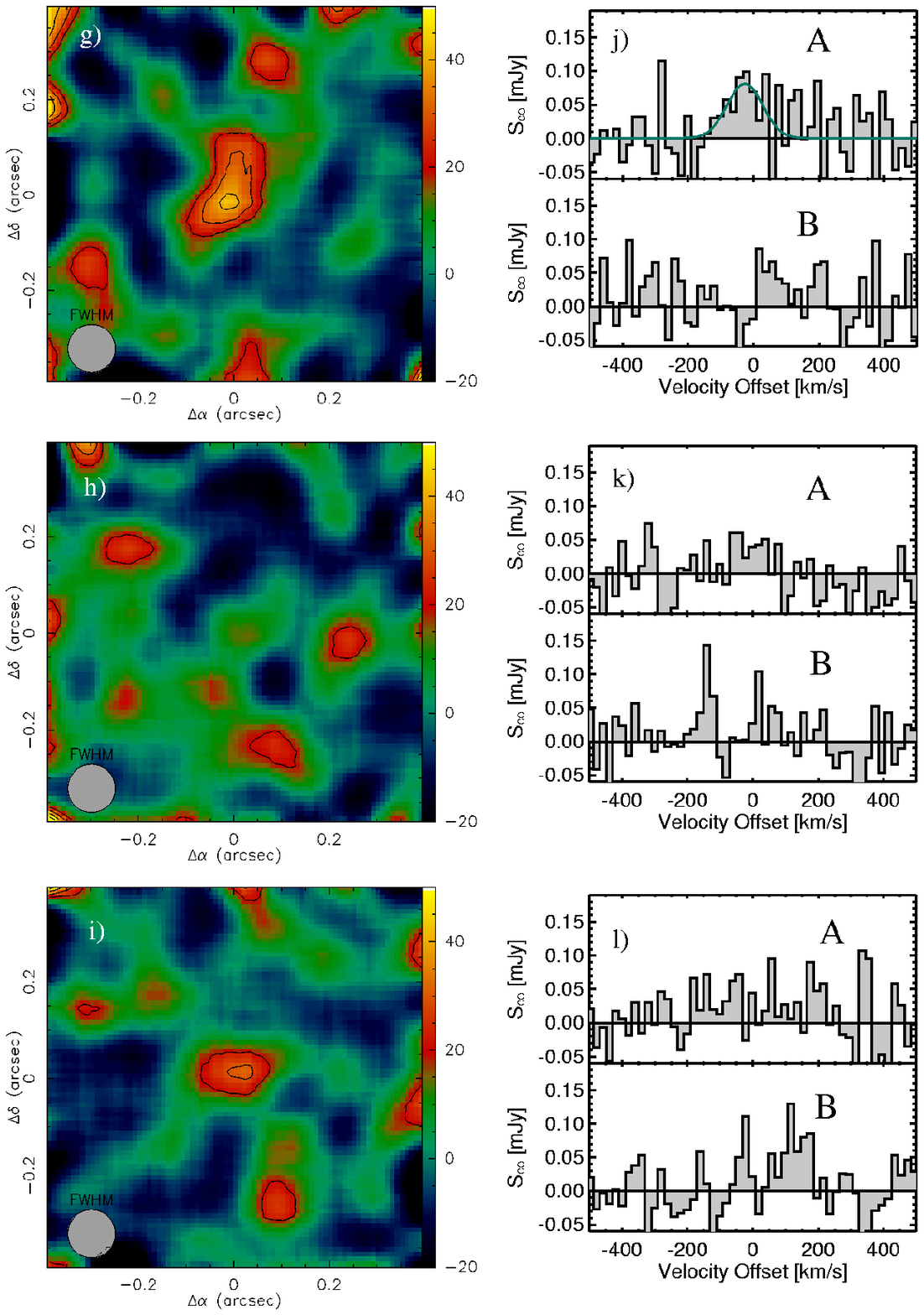}
      \caption{{\bf a-c):} Velocity-integrated CO 4--3 intensity maps (in units $\mu$Jy\,km\,s$^{-1}$) 
of our simulation as seen from three different lines of sight. Each panel is 6\,kpc on a side, with a pixel size
of 10\,pc. {\bf d-f):} The corresponding CO spectra integrated over the entire field (black curve), 
the A component (red) and its companion B (yellow). {\bf g-i):} CO 4--3 maps (in units $\mu\mbox{Jy}\,\mbox{beam}^{-1}$) 
of our simulated system in a-c) after 24\,hrs of integration with ALMA at $0.1\arcsecs$ resolution. 
The maps were obtained by averaging the channel-maps from -200\,km\,s$^{-1}$ to 
200\,km\,s$^{-1}$. The contours are $-2, 2, 3, 4\,\sigma$, where $\sigma = 10\,\mu$Jy\,beam$^{-1}$, 
where negative contours are shown as dashed curves. {\bf j-l):} The corresponding spectra (binned to 20\,km\,s$^{-1}$)
extracted from the centroid position of each source.}
         \label{figure:co-maps}
   \end{figure*}

The resulting CO 4--3 intensity maps of the inner 
6\,kpc$\times$6\,kpc (corresponding to $0.4\arcsecs \times 0.4\arcsecs$ at $z=3.26$) viewed along the $x, y$ and $z$ lines of 
sight are shown in Fig.\ \ref{figure:co-maps}a-c.
The CO 4--3 line is the lowest
transition which will be accessible to ALMA (band 3) at $z=3.26$.
It is seen that the bulk of the CO emission comes from the two merging disk systems, 
reflecting the fact that the molecular gas
is primarily situated in these two systems, and reaches its highest
densities there ($n(\mbox{H}_2)\sim 10^{6}\,\mbox{cm}^{-3}$).
However, faint CO emission is also seen to originate from filamentary structures 
extending out from the main system. These are merger-induced tidal streams of 
diffuse ($n(\mbox{H}_2)\sim 10^{1-2}\,\mbox{cm}^{-3}$) molecular gas, hence
the low CO surface brightness. The diffuse gas accounts for about 1 and 10 per-cent of
the total CO 4--3 and 1--0 luminosities, respectively. We find a global CO 4--3/1--0 line luminosity ratio of $\sim 0.6$ 
from our simulation, which is marginally subthermal. This, however, is merely an
'aperture effect', due to the fact that the extended, diffuse gas contributes
more strongly to the CO 1--0 than the 4--3 line emission.
The CO emission shows significant substructure within each of
the merging galaxies. In the face-on view of the central galaxy (Fig.\ \ref{figure:co-maps}a), the
ISM is seen to consist of multiple cavities of low CO emission, surrounded by
filamentary regions where the CO emission is strong. 
These structures are created by SNe explosions, which 
carve out 'holes' in the ISM, while at the same time
creating regions of dense gas as the blastwave ploughs through the surrounding medium,
compressing it in the process. Thus in our simulation, SNe appear to have a significant
impact on the distribution of molecular gas within a galaxy, which is consistent with
the notion that star formation sites are born in the wake of SNe.

The CO spectra integrated over the entire region
are shown in black in Fig.\ \ref{figure:co-maps}d-f. The first thing
to notice is the strong dependence of the line shape on
the viewing angle. As expected, the spectrum in Fig.\ \ref{figure:co-maps}d
is dominated by a very narrow ($\Delta V_{\mbox{\tiny{FWHM}}}\simeq 50\,\mbox{km}\,\mbox{s}^{-1}$) 
and smooth single profile, which corresponds to the
main disk viewed face-on. 
In contrast, the spectra in Fig.\ \ref{figure:co-maps}e-f
have double-horn profiles, which is normally taken to 
indicate orbital motion in a disk or a merger. 
While such double-horn profiles observed towards local 
ULIRGs have been shown to be due to a compact circumnuclear disk of molecular gas 
(e.g.\ Downes \& Solomon 1998), the issue is still
controversial for high-$z$ objects (Ivison et al.\ 2001; Downes \& Solomon 2003)
In our case, it is the merger-nature of the system which is responsible
for the overall double-horn profile. Although, we note that each disk system
(whose individual spectra are shown in red(A) and yellow(B)), when
viewed edge-on, also give rise to a double-peaked profile, albeit a less distinct one. 
The above illustrates that great care needs to be taken when interpreting high-$z$ observations,
and underlines the need for high-resolution observations capable of resolving the emission.
Finally, we note that the fact that the line profile in Fig.\ \ref{figure:co-maps}d 
is smooth, strongly suggest that while the molecular gas can be perturbed by SNe,
the velocities so attained and/or the amount of gas affected are insufficient to make
a significant imprint on the CO line profile. Of course this does not rule out the
possibility that in galaxies with more prodigal star formation rates, SNe, and possibly even stellar winds,
could have a significant effect on the observed CO line profile.  
Also, we note that Narayanan et al.\ (2006) found that feedback from an AGN in massive
merger systems can produce outflows of molecular gas, which are substantial enough
to be detectable in CO.

\subsection{Comparison with observations} 
Does the predicted CO properties of the simulated proto-galaxy resemble
those of observed $z\sim3$ LBGs?
Only two LBGs have been detected in CO to date: the 
gravitationally lensed systems MS\,1512$-$cB58 ($z=2.73$) and
J2135$-$0101 ($z=3.07$) (Baker et al.\ 2004; Coppin et al.\ 2007).
The two LBGs have star formation rates in the range $\sim 20-60\,\Msolar\,$yr$^{-1}$,
i.e.\ comparable to our simulated system.
The peak CO 3--2 line intensities towards cB58 and J2135 are $\sim (2\,$mJy$)/\mu$ and
$\sim (3\,$mJy$)/\mu$, where the lensing amplification factors are 
$\mu \simeq  31.8$ and $\simeq 8$, respectively.
Thus, their intrinsic CO line fluxes -- even if the
CO 4--3/3--2 brightness temperature ratios differ from unity
(Papadopoulos et al.\ 2000) -- are within the range predicted by our simulation.
While the intrinsic CO brightness of J2135 is $\sim 7\times$ 
that of cB58, Fig.\ \ref{figure:co-maps}d-f shows that such
differences can in principle come about merely from orientation effects. 
Integrating over the entire line, however, we find that the total CO
luminosity of our simulation is relatively constant (within $\sim 30$ per-cent)
with respect to inclination angle, suggesting that CO is a robust tracer of the gas mass.
The CO line widths of cB58 and J213 are very similar ($\Delta V_{\mbox{\tiny{FWHM}}}\simeq 170-190\,\mbox{km}\,\mbox{s}^{-1}$), 
and in excellent agreement with the edge-on CO line widths of components A and B. This
suggest that both systems are viewed edge-on (J2135 even has hint of a double-horn profile) and
that the difference in the CO properties of these two $z\sim 3$ LBGs is not due
to inclination effects, but due to 
an intrinsic diffference in the gas content of the two systems.

\subsection{CO as a dynamical mass tracer}
Mapping the CO emission from distant galaxies is one of the best ways
of determining their dynamical masses, although due to the inability of current 
interferometers to robustly resolve the CO emission in high-$z$ objects, most attempts
at deriving dynamical masses have resulted in little more than upper limits.
Here we aim to quantify just how well one can determine the dynamical
mass using CO. 

For each line of sight we have calculated the dynamical masses of A and B using
the disk-formula of Bryant \& Scoville (1996).
Comparing with their total masses ($M^A_{tot}=1.0\times 10^{10}\,\Msolar$ and $M^B_{tot}=0.5\times 10^{10}\,\Msolar$),
calculated as the sum of the total baryonic mass, i.e.\  stars and gas (atomic and molecular),
and DM, we find that in the best case scenario when the disks are seen edge-on, 
the dynamical masses can be inferred from the CO emission to within 20 per-cent. 
The virial mass of the overall merger-system is
given by $M_{vir} = 2RV^2/G$, where $R$ and $V^2$ are the mass-weighted
distance between the two systems and the
line-of-sight velocity variance.
The actual total mass of the entire system ($M_{tot}=2.0\times 10^{10}\,\Msolar$) 
is found by integrating up all the mass (baryonic and dark) within a sphere centered 
between the two systems and with a radius corresponding to half their distance.
We find that when the two systems are well-separated (Fig.\ \ref{figure:co-maps}b-c), the
total mass can be inferred to within 20 per-cent, whereas the discrepancy
increases to 50 per-cent when the two systems overlap as in Fig.\ \ref{figure:co-maps}a.

Observations of CO in distant galaxies have to deal with limited
sensitivity and spatial/spectral resolution, all of which can
affect dynamical mass estimates. How well can we expect to do such
observations using ALMA?
To answer this question we simulated a 24\,hr integration with ALMA of
our system (Fig.\ \ref{figure:co-maps}g-l). We adopted the projected rms noise of ALMA at $\sim 108\,$GHz
given by the online ALMA sensitivity calculator 
with default settings. The noise was added to each 'velocity-slice' in the CO 'data-cube', and convolved with a {\sc fwhm}~$=0.1\arcsecs$
Gaussian beam -- a resolution obtainable with ALMA in its extended configuration. 
From Fig.\ \ref{figure:co-maps}g-l it is seen that
only in the case where the system is viewed face-on
and is at its brightest, do we see a line profile and a 
detection ($4\sigma$) in the velocity-averaged maps.
Thus if our simulated system is representative of LBGs at $z\gs 3$, then ALMA observations 
of CO line emission from such systems will a) depend heavily on the inclination angle under
which they are viewed, thereby potentially bias detections towards face-on systems with
relatively small apparent CO linewidths, and b) essentially be pure detection experiments
since ALMA will not be sensitive enough to constrain CO sizes and line shapes.
The implication of a) and b) is that it may prove difficult to infer gas kinematics and
dynamical masses of sub-$L^*$ galaxies (such as the one simulated in this paper) at high-$z$
using ALMA.

\begin{acknowledgements}
\end{acknowledgements}

We thank Min Yang and Desika Narayanan for helpful discussions and the referee for improving the paper significantly.
The TreeSPH simulation was performed on the SGI Itanium II facility provided by DCSC.
The Dark Cosmology Centre is funded by the DNRF.


\begin{thebibliography}{}
\bibitem[Aalto et al.\ 1995]{Aalto-et-al-1995} Aalto S., et al.\ 1995, \aap, 300, 369.
\bibitem[Baker et al.\ 2004a]{Baker-et-al-2004a} Baker A.\ J., et al.\ 2004a, \apj, 604, 125.
\bibitem[Baker et al.\ 2004b]{Baker-et-al-2004b} Baker A.\ J., et al.\ 2004b, \apj, 613, L113.
\bibitem[Blain et al.\ (2002)]{Blain-et-al-2002} Blain A.\ W., et al.\ 2002, \physrep, 369, 111.
\bibitem[Brown and Vanden Bout (1992)]{Brown-and-Vanden-Bout-1992} Brown R.\ L.\ \& Vanden Bout P.\ A.\ 1992, \aj, 397, L19.
\bibitem[Bryant \& Scoville (1996)]{Bryant-and-Scoville-1996} Bryant P.\ M.\ \& Scoville, N.\ Z.\ 1996, \apj, 457, 678.
\bibitem[Cazaux \& Spaans (2004)]{Cazaux-Spaans-2004} Cazaux S.\ \& Spaans M.\ 2004, \apj, 611, 40.
\bibitem[Combes, Maoli \& Omont (1999)]{Combes-Maoli-and-Combes-1999} Combes F., Maoli R.\ \& Omont A.\ 1999, \aap, 345, 369.
\bibitem[Coppin et al.\ (2007)]{Coppin-et-al-2007} Coppin K.E.K., et al.\ 2007, ApJ, 665, 936.
\bibitem[Downes \& Solomon (1998)]{Downes-and-Solomon-1998} Downes D.\ \& Solomon P.\ M.\ 1998, \apj, 507, 615.
\bibitem[Downes \& Solomon (2003)]{Downes-and-Solomon-2003} Downes D.\ \& Solomon P.\ M.\ 2003, \apj, 582, 37.
\bibitem[Draine \& Salpeter (1978)]{Draine-and-Salpeter-1978} Draine B.\ T.\ \& Salpeter E.\ E.\ 1978, \nat, 271, 730. 
\bibitem[Elmegreen (1989)]{Elmegreen-1989} Elmegreen B.\ G.\ 1989, \apj, 338, 178. 
\bibitem[Elmegreen (1993)]{Elmegreen-1993} Elmegreen B.\ G.\ 1993, \apj, 411, 170.
\bibitem[Giavalisco (2002)]{Giavalisco-2002} Giavalisco M.\ 2002, \araa, 40 579.
\bibitem[Goldreich \& Kwan 1974]{Goldreich-and-Kwan-1974} Goldreich P.\ \& Kwan J.\ 1974, \apj, 189, 441.
\bibitem[Habing (1968)]{Habing-1968} Habing H.\ J.\ 1968, \bain, 19, 421.
\bibitem[Ivison et al.\ (2001)]{Ivison-et-al-2001} Ivison R.\ J., et al.\ 2001, \apj, 561, L45.
\bibitem[Larson (1981)]{Larson-1981} Larson R.\ B.\ 1981, \mnras, 194, 809.
\bibitem[Mirabel \& Sanders (1989)]{Mirabel-and-Sanders-1989} Mirabel I.\ F.\ \& Sanders D.\ B.\ 1989, \apj, 340, L53. 
\bibitem[Narayanan et al.\ (2006)]{Narayanan-et-al-2006} Narayanan D., et al.\ 2006, \apj, 642, L107.
\bibitem[Papadopoulos et al.\ (2000)]{Papadopoulos-et-al-2000} Papadopoulos P.\ P., et al.\ 2000, \apj, 528, 626.  
\bibitem[Pelupessy, Papadopoulos \& van der Werf (2006)]{Pelupessy-et-al-2006} Pelupessy I., et al.\ 2006, \apj, 645, 1024. 
\bibitem[Shapley et al.\ 2005]{Shapley-et-al-2005} Shapley A.\ E., et al.\ 2005, \apj, 626, 698.
\bibitem[Silk \& Spaans (1997)]{Silk-and-Spaans-1997} Silk J.\ \& Spaans M.\ 1997, \apj, 488, L79.
\bibitem[Solomon \& Vanden Bout (2005)]{Solomon-and-Vanden-Bout-2005} Solomon, P.\ \& Vanden Bout, P.\ 2005, \araa, 43, 677.
\bibitem[Sommer-Larsen (2006)]{Sommer-Larsen-2006} Sommer-Larsen J.\ 2006, \apj, 644, L1.
\bibitem[Sommer-Larsen (2007)]{Sommer-Larsen-2007} Sommer-Larsen J.\ \& Fynbo J.P.U.\ 2007, \mnras, submitted. 
\bibitem[Spergel et al.\ (2003)]{Spergel-et-al-2003} Spergel D.\ N., et al.\ 2003, \apjs, 148, 175.
\bibitem[Stark \& Blitz (1978)]{Stark-and-Blitz-1978} Stark A.\ A.\ \& Blitz L.\ 1978, \apj, 225, L15.
\end{thebibliography}
\end{document}